\documentclass[preprint,12pt]{elsarticle}

\usepackage{amssymb}
\usepackage{amsthm}

\journal{Physics Letter B}

\begin{document}

\begin{frontmatter}

\title{Justifying the Naive Partonic Sum Rule for Proton Spin}

\author[address1,address2,address3]{Xiangdong Ji}

\author[address1]{Jian-Hui Zhang}

\author[address3]{Yong Zhao}

\address[address1]{INPAC, Department of Physics and Astronomy, Shanghai Jiao Tong University, Shanghai, 200240, P. R. China}
\address[address2]{Center for High-Energy Physics, Peking University, Beijing, 100080, P. R. China}
\address[address3]{Maryland Center for Fundamental Physics, University of Maryland, College Park, Maryland 20742, USA}



\begin{abstract}
We provide a theoretical basis for understanding the spin structure of the proton in terms of the spin and orbital angular momenta 
of free quarks and gluons in Feynman's parton picture. We show that each term in the Jaffe-Manohar spin sum rule can be related to the matrix element of a
gauge-invariant, but frame-dependent operator through a matching formula in large-momentum effective field theory. We present all the matching 
conditions for the spin content at one-loop order in perturbation theory, which provide a basis to calculate parton orbital angular momentum 
in lattice QCD at leading logarithmic accuracy.
\end{abstract}


\end{frontmatter}

Understanding the spin structure of the proton has been an important goal in hadron physics. In the past 25 years, two well-known spin sum rules have been proposed to analyze the content of the proton spin. The first, proposed by Jaffe and Manohar~\cite{Jaffe:1989jz}, was motivated from a free-field expression of QCD angular momentum boosted to the infinite momentum frame (IMF) of the proton. The second is the frame-independent and manifestly gauge-invariant decomposition by one of the authors~\cite{Ji:1996ek}. Notwithstanding that the latter has received considerable attention for its relation to generalized parton distributions (GPDs) and experimental probes~\cite{Ji:1996ek, Ji:1996nm, Hoodbhoy:1998yb}, it is not natural in the 
language of parton physics (see, however, a recent discussion on its connection to the transverse polarization~\cite{Ji:2012vj}). In contrast, the Jaffe-Manohar sum rule defined in the light-cone gauge $A^+=0$ has a natural partonic interpretation, as the proton spin can be decomposed into four parts,
\begin{equation}
    \frac{1}{2}= \frac{1}{2}\Delta \Sigma(\mu) + \Delta L_q(\mu) + \Delta G(\mu) + \Delta L_g(\mu) \ ,
\end{equation}
where the individual terms are the spin and orbital angular momenta (OAM) of the quark and gluon partons, respectively, and $\mu$ is the renormalization scale. All the four terms are defined to be the proton matrix elements of free-field angular momentum operators (AMOs) in IMF~\cite{Jaffe:1989jz}:
\begin{eqnarray}
\vec{J} &=& \int d^3x\ \psi^\dagger \frac{\vec{\Sigma}}{2} \psi + \int d^3x\ \psi^\dagger\vec{x}\times (-i\vec{\nabla} )\psi \nonumber\\
&&+ \int d^3x\ \vec{E}_a\times\vec{A}^a + \int d^3x\ E_a^i\ \vec{x}\times\vec{\nabla} A^{i,a} \ ,
\label{jm}
\end{eqnarray}
where $E^i=F^{i+}$, $a$ and $i$ are the color and spatial indices. However, the free-field form of the angular momentum in gauge theories faces a conceptual problem: 
all terms except the first one are gauge dependent, and it is unclear why the light-cone gauge operator is measurable in physical experiments.

In the past two decades, there has been a long list of literatures attempting to justify
the Jaffe-Manohar sum rule as physical (see e.g.,~\cite{Jaffe:1995an, Chen:1998iu,Leader:2011za,Hoodbhoy:1999dr,Ji:2013fga,Ji:2012gc,Hatta:2013gta}). There are strong motivations behind this: First, $\Delta G$ as defined in the light-cone gauge is measurable 
in high-energy experiments, although this appears to be a theoretical puzzle by itself---while $\Delta G$ is easy to define from the Feynman parton
picture, there is no natural gauge-invariant notion for the spin of gauge particles~\cite{landau}. Second,
when the proton is probed in IMF, some of its physical properties can be understood from simple addition of those of free quarks and gluons. For example, the longitudinal momentum of the proton is the sum of that of the quark and gluon partons:
\begin{equation}
                1=\int dx\ x\left(\sum_q q(x) + g(x)\right)  \ ,
\end{equation}
where $q(x)$ and $g(x)$ are the unpolarized quark and gluon momentum distribution functions. This simple parton picture may work for the proton spin as well.

It was first proposed that although the free-field AMOs are gauge dependent, their physical matrix elements are gauge invariant~\cite{Chen:1998iu}. A similar claim was also made recently~\cite{Leader:2011za}. However, this is invalidated by a one-loop calculation by the present authors~\cite{Ji:2013fga}, where the matrix element of the gluon spin operator was shown to be different in the Coulomb and light-cone gauges~\cite{Ji:1995cu} (for more general discussions see Ref.~\cite{Hoodbhoy:1999dr}). Actually, as argued in Refs.~\cite{Ji:2012gc, Ji:2013fga}, for the bound-state proton, there is no physically meaningful notion of gluon spin or OAM due to the existence of longitudinal gluons. Only when the proton is boosted to IMF, the longitudinal component of gluons is suppressed by the infinite boost and the gluons can be regarded as free radiation. This is the well-known Weizs\"{a}cker-Williams (WW) approximation~\cite{Jackson}. The gluon spin then acquires a clear physical meaning and can be represented by $\vec{E}\times\vec{A}$, but is subject to a class of ``physical" gauge conditions that leave the transverse polarizations of the gluon field intact~\cite{Hatta:2013gta}. Similar arguments also apply to the quark and gluon OAM. Therefore, 
we can regard the free-field form in the Jaffe-Manohar sum rule as physical if we work in IMF with a ``physical" gauge condition. This is equivalent to using the light-cone coordinates and gauge~\cite{Jaffe:1989jz}, and the reason is simple: All the ``physical" gauges will flow into the light-cone gauge in the IMF limit~\cite{Hatta:2013gta}.

From a practical perspective, the Jaffe-Manohar sum rule still poses difficulty for a nonperturbative lattice calculation of its individual contributions, 
because the explicit usage of light-cone coordinates and gauge brings real-time dependence. One may avoid this difficulty by using normal space-time coordinates with 
a ``physical" gauge that does not involve time, and calculating with a proton at infinite momentum. However, the largest momentum attainable on the lattice with spacing $a$ is 
constrained by the lattice cutoff $\pi/a$.

The above difficulty can, however, be circumvented in the framework of {\it large-momentum effective field theory} (LaMET)~\cite{Ji:2014gla} proposed by one of the authors. Suppose one is to calculate some light-cone or parton observable $\mathcal O$. Instead of computing it directly, one defines, in the LaMET framework, a quasi-observable $\tilde{\mathcal{O}}$ that depends on a large hadron momentum $P^z$. In general, both $\mathcal O$ and $\tilde{\mathcal{O}}$ suffer from ultraviolet (UV) divergences. If $P^z\to\infty$ is taken prior to a UV regularization, the quasi-observable $\tilde{\mathcal O}$ becomes the parton observable $\mathcal{O}$ by construction. However, what one can calculate in practice is the quasi-observable $\tilde{\mathcal O}$ at large but finite $P^z$ with UV regularization imposed first. This is the case in lattice computations. The difference between $\mathcal O$ and $\tilde{\mathcal O}$ is just the order of limits. This is similar to an effective field theory set-up. The difference is that here the role of perturbative degrees of freedom is played by the large momentum of the external state, hence it cannot be arranged into a Lagrangian formalism. Nevertheless, one can bridge the quasi- and parton observables through 
\begin{equation}
\tilde{\mathcal{O}}(P^z/\Lambda) = Z\left(P^z/\Lambda, \mu/\Lambda\right) \mathcal{O}(\mu) + {c_2\over (P^z)^2} + {c_4\over (P^z)^4} + \cdots \ ,
\label{eft}
\end{equation}
where $\Lambda$ is a UV cutoff imposed on the quasi-observable, and $c_i$'s are higher-twist contributions suppressed by powers of $P^z$. That is, the quasi-observable $\tilde{\mathcal{O}}(P^z/\Lambda)$ can be factorized into the parton observable $\mathcal{O}(\mu)$ and a matching coefficient $Z$, up to power suppressed corrections. Taking the $P^z\to\infty$ limit does not change the infrared (IR) behavior of the observable, but only affects its UV behavior. Therefore $\mathcal{O}(\mu)$ captures all the IR physics in $\tilde{\mathcal{O}}(P^z/\Lambda)$, and the matching coefficient $Z$ is completely perturbative.

An explicit example of Eq.~(\ref{eft}) is presented in Refs.~\cite{Ji:2013dva,Xiong:2013bka} for the case of parton distribution functions (PDFs), where the factorization formula has a convolution form, and the matching coefficients were calculated at the leading logarithmic order. Using these results, the first direct lattice calculation of the isovector sea-quark parton distributions has been available recently~\cite{Lin:2014zya}. A similar factorization formula was also proposed in Ref.~\cite{Ma:2014jla} to extract PDFs from lattice QCD calculations based on QCD factorization of lattice ``cross sections".

Within the LaMET framework, we can start with suitable quasi-observables to calculate the proton spin content. According to our discussions above, these quasi-observables can be defined as the free-field QCD AMOs in a ``physical" gauge condition that has the correct WW approximation in the IMF limit~\cite{Hatta:2013gta}. A possible choice of the ``physical" gauge condition is the expression in terms of nonlocal operators introduced by Chen {\it et al.}~\cite{Chen:2008ag, Chen:2009mr}:
\begin{eqnarray}
   \vec{J}_{\rm QCD} &= \int d^3x\ \psi^\dagger \frac{\vec{\Sigma}}{2} \psi
   + \int d^3x\ \psi^\dagger \vec{x}\times(-i\vec{\nabla} - g\vec{{A}}_\parallel)\psi\nonumber\\
&+ \int d^3x\ \vec{E}_a\times\vec{A}^a_\perp + \int d^3x\ E_a^i\ (\vec{x}\times\vec{\nabla}) A^{i,a}_\perp \ ,
\label{chen}
\end{eqnarray}
where $\vec{x}$ are the spatial coordinates, and $\vec{A}$ is decomposed into a pure-gauge part $\vec{A}_\parallel$ and a physical part $\vec{A}_\perp$ which satisfy (see also Ref.~\cite{Treat:1973yc})
\begin{eqnarray}
     \partial^i A^{j,a}_{\parallel} - \partial^j A^{i,a}_{\parallel} -gf^{abc}A_{\parallel}^{i,b}A_{\parallel}^{j,c} &=& 0 \ ,\nonumber\\
\label{nullfieldstrength}
\partial^i {A}^i_\perp - ig[{A}^i, {A}^i_\perp] &=& 0\ ,
\label{coulomb}
\end{eqnarray}
so that each term in Eq.~(\ref{chen}) is gauge invariant. From Eq.~(\ref{coulomb}), one can show that in the Coulomb gauge $\vec{\nabla}\cdot\vec{A}=0$, $\vec A_\perp$ equals $\vec A$ order by order in perturbation theory. Therefore, Eq.~(\ref{chen}) corresponds to choosing the Coulomb gauge as the ``physical" gauge.


It has been shown in Ref.~\cite{Ji:2013fga} that $\vec{E}_a\times \vec{A}^a_\perp$ in Eq.~(\ref{chen}) is equivalent to the total gluon spin operator in the IMF limit. It is easy to see that the other nonlocal terms in Eq.~(\ref{chen}) also have the correct WW approximation as the parton OAM. 
Therefore, we can choose the nonlocal operators in Eq.~(\ref{chen}) as the quasi-observables for parton angular momentum in the LaMET approach.



The advantage of the expression in Eq.~(\ref{chen}) is that it is time independent and thus allows for a direct calculation in lattice QCD. Suppose we evaluate the matrix elements of these quasi-observables with finite momentum $P^z$, we should have
\begin{equation}
{1\over2} = {1\over2}\Delta\widetilde\Sigma(\mu, P^z) + \Delta \widetilde G(\mu, P^z) + \Delta \widetilde L_q(\mu, P^z) + \Delta \widetilde L_g(\mu, P^z)\ ,
\end{equation}
where the $P^z$-dependence is expected since Eq.~(\ref{chen}) is a frame-dependent expression~\cite{Ji:2013fga}. Following the effective theory argument above, we can relate these quasi-observables to the corresponding parton observables through the following factorization formula:
\begin{eqnarray}
\Delta\widetilde\Sigma(\mu, P^z) &=& \Delta \Sigma(\mu) \ ,\nonumber\\
\Delta\widetilde G(\mu, P^z) &=& z_{qg} \Delta \Sigma(\mu) + z_{gg}\Delta G(\mu) +O\left({M^2\over (P^z)^2}\right)\ ,\nonumber\\
\Delta\widetilde L_q(\mu, P^z) &=& P_{qq} \Delta L_q(\mu) + P_{gq}\Delta L_g(\mu)\nonumber\\
&& + p_{qq} \Delta \Sigma(\mu) + p_{gq}\Delta G(\mu) + O\left({M^2\over (P^z)^2}\right) \ ,\nonumber\\
\Delta\widetilde L_g(\mu, P^z) &=& P_{qg} \Delta L_q(\mu) + P_{gg}\Delta L_g(\mu)\nonumber\\
&& + p_{qg} \Delta \Sigma(\mu)  + p_{gg}\Delta G(\mu) +O\left({M^2\over (P^z)^2}\right) \ ,
\label{factorization}
\end{eqnarray}
where $M$ is the proton mass, and all the matrix elements are renormalized in dimensional regularization and $\overline{\mbox{MS}}$ scheme. 
$\Delta\widetilde\Sigma(\mu, P^z)$ is the same as $\Delta \Sigma(\mu)$ because the quark spin operator is frame independent and should have the same matrix elements in the Coulomb and light-cone gauges. The $z_{ij}$, $P_{ij}$ and $p_{ij}$'s are the matching coefficients to be calculated in perturbative QCD.

In the remainder of this paper, we show how to obtain all the matching coefficients in Eq.~(\ref{factorization}) at one-loop order. First, let us take $z_{qg}$ and $z_{gg}$ as an example. To obtain $z_{qg}$ and $z_{gg}$, we need to calculate the matrix elements of $\vec{E}_a\times\vec{A}^a_\perp$ at finite $P^z$ and in the IMF limit (before UV regularization). To ensure gauge invariance and angular momentum conservation in our calculation, we use on-shell and massless external quarks and gluons, and regularize the UV and IR/collinear divergences with dimensional regularization ($d=4-2\epsilon$). One may think of using the off-shellness of external quarks and gluons as IR/collinear regulator, and then take the on-shell limit. However, in this case one needs to take into account the contribution from the ghost and gauge-fixing terms. This is because the total angular momentum operator in QCD from Noether's theorem contains not only the terms presented in our paper, but also the ghost and gauge-fixing terms---which are called BRS-exact ``alien" operators in Ref.~\cite{Collins:1994ee}---from the QCD Lagrangian. The matrix elements of BRS-exact operators vanish in a physical on-shell state, but not in an off-shell state. Therefore, one has to be careful with these contributions when starting from off-shell external states and then going to the on-shell limit, in order to have angular momentum conservation. Considering matrix elements of on-shell states simply avoids such complications.

Since the AMOs we consider are all gauge invariant, we can work in any gauge, and for simplicity we choose the Coulomb gauge. 
As mentioned before, the Coulomb gauge condition is equivalent to the condition for $\vec{A}_\perp$ in QCD (see Eq.~(\ref{coulomb})).

At tree level, $\Delta\widetilde G^{\scriptsize\mbox{tree}}=\Delta G^{\scriptsize\mbox{tree}}$. At one-loop level, the matrix element of $\vec{E}_a\times\vec{A}^a_\perp$ is

\begin{eqnarray}
\Delta\widetilde G^{(1)} &=& {\alpha_SC_F\over 4\pi} \left({5\over3}{1\over\epsilon'_{UV}} + {4\over3}\ln{(P^z)^2\over\mu^2} - {3\over\epsilon'_{IR}} +R_1\right)\Delta \Sigma^{\scriptsize\mbox{tree}} \nonumber\\
&&+{\alpha_S\over 4\pi} \left[{4C_A- 2n_f\over3}{1\over\epsilon'_{UV}} - {11C_A- 2n_f\over3}{1\over\epsilon'_{IR}} + C_A\left({7\over3}\ln {(P^z)^2\over\mu^2}+R_2\right) \right]\Delta G^{\scriptsize\mbox{tree}}\ ,\nonumber\\
\end{eqnarray}
where $n_f$ is the number of flavors of active quarks, and
\begin{equation}
{1\over\epsilon'} = {1\over\epsilon} - \gamma_E + \ln 4\pi\ , \ \ C_F = {N_c^2 -1 \over 2N_c} \ , \ \ C_A = N_c\ ,
\end{equation}
with $N_c$ being the number of colors. $R_1$ and $R_2$ are finite constants
\begin{equation}
R_1 \ = \ {8\over3}\ln2 - {64\over9}, \ \ \ \ \ \ \ \ R_2\ = \ {14\over3}\ln2 - {121\over9}\ .
\end{equation}

The corresponding IMF (or light-cone) matrix elements are~\cite{Ji:1995cu}
\begin{eqnarray}
\Delta G^{(1)} &=& {\alpha_SC_F\over 4\pi} \left({3\over\epsilon'_{UV}} - {3\over\epsilon'_{IR}}\right)\Delta \Sigma^{\scriptsize\mbox{tree}} \nonumber\\
&&+ {\alpha_S\over 4\pi} \left[{11C_A- 2n_f\over3}\left({1\over\epsilon'_{UV}} - {1\over\epsilon'_{IR}}\right)\right]\Delta  G^{\scriptsize\mbox{tree}}\ .\nonumber\\
\end{eqnarray}

Apparently the anomalous dimensions (coefficients of $1/\epsilon'_{UV}$'s) are different between $\Delta\widetilde G^{(1)}$ and $\Delta G^{(1)}$, but the IR divergence (coefficients of $1/\epsilon'_{IR}$'s) is the same for both. After renormalization, we substitute the $1/\epsilon'_{IR}$ terms in $\Delta\widetilde G$ with $\Delta G$,
and obtain at $O(\alpha_S)$
\begin{eqnarray}
\Delta\widetilde G &=& {\alpha_SC_F\over4\pi}\left({4\over3}\ln{(P^z)^2\over \mu^2}+R_1\right) \Delta\Sigma \nonumber\\
&&+ \left[1+{\alpha_SC_A\over4\pi}\left({7\over3}\ln{(P^z)^2\over \mu^2}+R_2\right)\right] \Delta G\ .
\end{eqnarray}

Therefore, the matching coefficients for $\Delta\widetilde G$ can be read off as
\begin{eqnarray}
z_{qg}(\mu/P^z) &=& {\alpha_SC_F\over4\pi} \left({4\over3}\ln{(P^z)^2\over \mu^2}+R_1\right)
\ ,\nonumber\\
z_{gg}(\mu/P^z) &=& 1+ {\alpha_SC_A\over4\pi} \left({7\over3}\ln{(P^z)^2\over \mu^2} +R_2\right) \ .
\end{eqnarray}

Following the same procedure, we can calculate all the other matching coefficients in Eq.~(\ref{factorization}) at one-loop order. The complete results are as follows:
\begin{equation}
\begin{array}{lrl}
\displaystyle P_{qq} =\ 1+ {\alpha_SC_F\over4\pi} \left(-2\ln{(P^z)^2\over\mu^2} +R_3 \right)  &, &\displaystyle P_{gq} \ =\ 0\ ,\\
\displaystyle P_{qg} =\ {\alpha_SC_F\over4\pi} \left(2\ln{(P^z)^2\over\mu^2}-R_3\right)  &, &\displaystyle P_{gg} \ =\ 1\ ,\\
\displaystyle p_{qq} =\  {\alpha_SC_F\over4\pi} \left(-{1\over3}\ln{(P^z)^2\over\mu^2} +R_4 \right) &, &\displaystyle p_{gq} \ =\ 0\ ,\\
\displaystyle p_{qg} =\ {\alpha_SC_F\over4\pi} \left(-\ln{(P^z)^2\over\mu^2}-R_1-R_4\right)  &, &\\
\displaystyle p_{gg} \ =\ {\alpha_SC_A\over4\pi} \left(-{7\over3}\ln{(P^z)^2\over\mu^2} -R_2 \right) &,&
\end{array}
\label{leadinglog}
\end{equation}
where
\begin{equation}
R_3 \ = \ -4\ln2 + {28\over3} \ ,\ \ \ \ \ \ \ R_4\ =\ -{2\over3}\ln2 + {13\over9} \ .
\end{equation}

With the above results, we are able to convert the quasi-observables in Eq.~(\ref{chen}) evaluated at a large finite momentum to the parton spin and OAM in IMF. This can be done by a simple inversion of Eq.~(\ref{factorization}). Although for realistic lattice computations the above matching coefficients have to be recalculated using lattice perturbation theory~\cite{Capitani:2002mp}, the leading logarithmic term of the matching coefficients is independent of the regularization scheme, and therefore is the same in dimensional and lattice regularizations. Our one-loop matching coefficients can thus be used for an approximate lattice computation of parton angular momentum to leading logarithmic accuracy.


In summary, we have justified the physical significance of the Jaffe-Manohar spin sum rule as a result of the WW approximation in IMF. In addition, we have shown how to obtain the partonic contributions to proton spin using the LaMET approach. The solution is a perturbative factorization formula that allows us to extract the parton spin and OAM in IMF from lattice QCD calculations with a finite but large proton momentum. Since the OAM in the Jaffe-Manohar sum rule can be related to experimentally measurable distributions such as twist-2 and -3 GPDs~\cite{Hatta:2011ku,Ji:2012sj}, we can eventually compare the parton OAM in theory and experiment.

\vspace{2em}
We thank J.~-W. Chen for useful discussions about the gauge-invariant gluon spin operator. We also thank M.~Glatzmaier and K.~-F. Liu for discussions on matching in lattice perturbation theory. This work was partially supported by the U.S. Department of Energy Office of Science, Office of Nuclear Physics under Award Number DE-FG02-93ER-40762
and a grant (No. 11DZ2260700) from the Office of
Science and Technology in Shanghai Municipal Government,  and grants by the National Science Foundation of China (No. 11175114, No. 11405104).


\end{document}